\documentclass[10pt,reqno]{article}
\usepackage{fullpage} 
\usepackage{amssymb} 
\usepackage{graphicx} 
\usepackage{amsmath} 
\usepackage{amsthm,amssymb,latexsym} 
\usepackage{amstext, amsfonts,amsbsy} 
\usepackage{enumerate} 
\usepackage{upgreek} 
\usepackage{float} 
\usepackage{color} 
\usepackage{accents} 
\usepackage{mathtools} 
\usepackage{tikz}
\usetikzlibrary{matrix,graphs,arrows,positioning,calc,decorations.markings,shapes.symbols,shapes.geometric,shapes.misc}
\usepackage{calc, ifthen, ae, xstring, etoolbox, xkeyval,pgfplots}
\usepackage[pdftex,bookmarks,colorlinks,breaklinks]{hyperref}  
\definecolor{dullmagenta}{rgb}{0.4,0,0.4}   
\definecolor{darkblue}{rgb}{0,0,0.4}
\hypersetup{linkcolor=red,citecolor=blue,filecolor=dullmagenta,urlcolor=darkblue} 
\usepackage{eucal}

\newtheorem{theorem}{Theorem}

\newtheorem{lemma}[theorem]{Lemma}

\theoremstyle{definition}

\theoremstyle{remark}

\newtheorem*{notation*}{Notation} 
\newtheorem*{acknowledgement}{Acknowledgements}

\numberwithin{equation}{section}

\begin{document}

{\noindent\Large\bf Orthogonal Polynomials for the Gaussian Weight with a Jump and Discrete Painlev\'e Equations 
}
\medskip

\begin{flushleft}

\textbf{Anton Dzhamay}\\
Beijing Institute of Mathematical Sciences and Applications (BIMSA)\\ 
No.~544, Hefangkou Village Huaibei Town, Huairou District, Beijing 101408 \\
E-mail: \href{mailto:adzham@bimsa.cn}{\texttt{adzham@bimsa.cn}}\qquad 
ORCID ID: \href{https://orcid.org/0000-0001-8400-2406}{\texttt{0000-0001-8400-2406}}\\[5pt]

\textbf{Elizaveta Trunina}\\
Department of Physics and Astronomy, Rutgers University, Piscataway, NJ 08854 USA\\
E-mail: \href{mailto:et438@physics.rutgers.edu}{\texttt{et438@physics.rutgers.edu}}\qquad

\emph{Keywords}: discrete Painlev\'e equations,  birational transformations, affine Weyl groups, Cremona transformations\\[3pt]

\emph{MSC2020}: 33D45, 34M55, 34M56, 14E07, 39A13

\end{flushleft}

%
%
\begin{abstract}
We describe a refined version of the \emph{discrete Painlev\'e identification problem} that emphasizes the importance on going beyond
just the surface type in describing a discrete Painlev\'e dynamic. We give an example of solving such identification problem for a particular recurrence
relation that previously  appeared in studying orthogonal polynomials for the Gaussian weight with a single jump.
\end{abstract}

\section{Introduction} 
\label{sec:intro}
By now it is clear that in studying various families of orthogonal polynomials one often encounters examples of non-linear second order
recurrence relations that turn out to be \emph{discrete Painlev\'e equations} in disguise. This is very similar to a better known continuous 
case and a very nice introduction into this topic is given in the recent book by W.~Van Assche \cite{Van:2018:OPPE}. Then a natural question 
is how to identify such recurrence relations as discrete Painlev\'e equations and also determine their type according to some classification 
scheme, which is a discrete analogue of what is sometimes called the \emph{Painlev\'e equivalence problem} \cite{Cla:2019:OPPE}. The classification 
scheme for both continuous and discrete Painlev\'e equations is due to H.~Sakai \cite{Sak:2001:RSAWARSGPE} and it is based on the geometric 
theory of Painlev\'e equations, see an excellent survey paper \cite{KajNouYam:2017:GAPE}. In a series of recent papers, e.g., \cite{DzhFilSto:2020:RCDOPWHWDPE} and \cite{DzhFilLigSto:2024:DHDPETIUGA},
it was shown that the geometric theory is also a very effective tool for solving the 
\emph{Painlev\'e identification problem}, since there exists an algorithmic scheme for reducing both discrete and continuous Painlev\'e equations
to their standard forms as listed in \cite{Sak:2001:RSAWARSGPE}, \cite{KajNouYam:2017:GAPE}, \cite{Oka:1980:PHAWPE}. 

Sakai's classification gives $22$ classes of different \emph{surface types} that appear as configuration spaces for the 
Painlev\'e dynamics. However, it is important to point out that this classification can in fact be refined -- 
it is not enough to determine the \emph{surface type} of the discrete Painlev\'e recurrence, 
since there are (infinitely many) non-equivalent discrete Painlev\'e dynamics that have the same surface type. 
In particular, it was shown in \cite{LiDzhFilZha:2025:RRGLCOPDPDS} that although recurrence relations that appear in studying orthogonal 
polynomials with generalized Laguerre and generalized Charlier weights both correspond to discrete Painlev\'e equations on the 
same $D_{6}^{(1)}$ surface, the dynamics they define are completely different and have different continuous limits. 

For recurrence relations that appear in studying orthogonal polynomials, another interesting question
is dependence of the weights on various parameters --- taking various parameter degenerations it is possible to transform one weight into 
another, and it is interesting to see if such degeneration goes through on the geometric level of Sakai surfaces. One recent example when 
this happens for the degeneration from the $q$-Racah to $q$-Hahn weight was considered in \cite{DzhKni:2020:QEQPDPE}

Therefore, we believe that it is important to better understand not just the geometry, but also the exact dynamics, 
as well as the effect of the initial conditions, for discrete Painlev\'e equations that occur in applications. 
For that we need to have a good collection of examples and in this paper we consider one such example -- 
the Gaussian weight with a single jump. We show that this example is related to the standard discrete
Painlev\'e equation on the $E_{6}^{(1)}$ Sakai surface with one additional geometry constraint that changes the 
symmetry group of the equation. We also compare this example with the related example of the modified Laguerre weight that
corresponds to the generic geometry.

%


\section{The Sakai Classification Scheme and Geometric Origins of Discrete Painlev\'e Equations} 
\label{sec:Sakai-theory}
We begin with a very brief introduction
into the geometric theory of Painlev\'e equations and explain precisely what we mean by the
\emph{Painlev\'e identification problem}. Almost all details have to be omitted and
we strongly recommend \cite{KajNouYam:2017:GAPE}, as well as the original paper \cite{Sak:2001:RSAWARSGPE},
for a careful introduction into this topic.

The definite classification scheme for Painlev\'e equations was suggested in the seminal paper by H.~Sakai \cite{Sak:2001:RSAWARSGPE}, who gave 
the complete classification of possible \emph{configuration spaces} for discrete Painlev\'e dynamics. Each such configuration space is a \emph{family}
of certain rational algebraic surfaces, known as \emph{generalized Halphen surfaces}. Recall that a generalized Halphen surface $\mathcal{X}$ is a
smooth projective rational surface that has an \emph{anti-canonical divisor of canonical type}, i.e., there exists an effective  
divisor $d\in |-\mathcal{K}_{X}|$, where $\mathcal{K}_{\mathcal{X}} = [K_{\mathcal{X}}]$ is the canonical class,  and if we consider 
its decomposition into irreducible components $d = \sum_{i} m_{i} d_{i}$ with multiplicities $m_{i}\in \mathbb{Z}_{>0}$, the intersection condition
$\mathcal{K}_{\mathcal{X}}\bullet [d_{i}] = d\bullet d_{i} = 0$ holds. The intersection configuration of the components $d_{i}$ can be encoded
by some \emph{affine} Dynkin diagram $\mathcal{R}$, whose type is called the \emph{surface type} of the equation. Sakai proved \cite{Sak:2001:RSAWARSGPE} that there are $22$ possible surface types that are shown on Figure~\ref{fig:Sakai-clsc-surf}. The upper indices 
$\left(\mathcal{R}\right)^{\text{e,q,d,c}}$ indicate the type of Painlev\'e dynamics on this configuration space --- \emph{(e)lliptic-difference},
\emph{multiplicative} (or \emph{($q$)-difference}), \emph{additive} (or \emph{(d)ifference}), and \emph{(c)ontinuous} (or \emph{differential}).
We call rational surfaces appearing in this classification scheme \emph{Sakai surfaces}, but it is important to point out that 
Sakai's approach generalized the earlier foundational work by K.~Okamoto in the continuous 
case, \cite{Oka:1979:FAESOPCFPP}, where such surfaces appeared as \emph{Okamoto spaces of initial conditions} for 
Hamiltonian forms of differential Painlev\'e equations, see Figure~\ref{fig:Sakai-clsc-surf}. 
\begin{center}
\begin{figure}[h]{\small
	\begin{tikzpicture}[>=stealth,scale=0.9]
		\node (e8e) at (2,4) {$\left(A_{0}^{(1)}\right)^{\text{e}}$};
		\node (a1qa) at (16,4) {$\left(A_{7}^{(1)}\right)^{\text{q}}$};
		\node (e8q) at (2,2) {$\left(A_{0}^{(1)*}\right)^{\text{q}}$};
		\node (e7q) at (4,2) {$\left(A_{1}^{(1)}\right)^{\text{q}}$};
		\node (e6q) at (6,2) {$\left(A_{2}^{(1)}\right)^{\text{q}}$};
		\node (d5q) at (8,2) {$\left(A_{3}^{(1)}\right)^{\text{q}}$};
		\node (a4q) at (10,2) {$\left(A_{4}^{(1)}\right)^{\text{q}}$};
		\node (a21q) at (12,2) {$\left(A_{5}^{(1)}\right)^{\text{q}}$};
		\node (a11q) at (14,2) {$\left(A_{6}^{(1)}\right)^{\text{q}}$};
		\node (a1q) at (16,2) {$\left(A_{7}^{(1)}\right)^{\text{q}}$};
		\node (a0q) at (18,2) {$\left(A_{8}^{(1)}\right)^{\text{q}}$};
		\node (e8d) at (4,0) {$\left(A_{0}^{(1)**}\right)^{\text{d}}$};
		\node (e7d) at (6,0) {$\left(A_{1}^{(1)*}\right)^{\text{d}}$};
		\node (e6d) at (8,0) {$\left(A_{2}^{(1)*}\right)^{\text{d}}$};
		\node (d4d) at (10,0) {$\left(D_{4}^{(1)}\right)^{\text{d,c}}$};
		\node (a3d) at (12,0) {$\left(D_{5}^{(1)}\right)^{\text{d,c}}$};
		\node (a11d) at (14,0) {$\left(D_{6}^{(1)}\right)^{\text{d,c}}$};
		\node (a1d) at (16,0) {$\left(D_{7}^{(1)}\right)^{\text{d,c}}$};
		\node (a0d) at (18,0) {$\left(D_{8}^{(1)}\right)^{\text{d,c}}$};
		\node (a2d) at (14,-2) {$\left(E_{6}^{(1)}\right)^{\text{d,c}}$};
		\node (a1da) at (16,-2) {$\left(E_{7}^{(1)}\right)^{\text{d,c}}$};
		\node (a0da) at (18,-2) {$\left(E_{8}^{(1)}\right)^{\text{c}}$};
		\draw[->] (e8e) -> (e8q);	\draw[->] (a1qa) -> (a0d);
		\draw[->] (e8q) -> (e7q); 	\draw[->] (e8q) -> (e8d);
		\draw[->] (e7q) -> (e6q); 	\draw[->] (e7q) -> (e7d);
		\draw[->] (e6q) -> (d5q); 	\draw[->] (e6q) -> (e6d);
		\draw[->] (d5q) -> (a4q); 	\draw[->] (d5q) -> (d4d);
		\draw[->] (a4q) -> (a21q); 	\draw[->] (a4q) -> (a3d);
		\draw[->] (a21q) -> (a11q); \draw[->] (a21q) -> (a11d); \draw[->] (a21q) -> (a2d);
		\draw[->] (a11q) -> (a1q); 	\draw[->] (a11q) -> (a1d); 	\draw[->] (a11q) -> (a1qa); 	\draw[->] (a11q) -> (a1da);
		\draw[->] (a1q) -> (a0q); 	\draw[->] (a1q) -> (a0d);	\draw[->] (a1q) -> (a0da);
		\draw[->] (e8d) -> (e7d);
		\draw[->] (e7d) -> (e6d);
		\draw[->] (e6d) -> (d4d);
		\draw[->] (d4d) -> (a3d);
		\draw[->] (a3d) -> (a11d);	\draw[->] (a3d) -> (a2d);
		\draw[->] (a11d) -> (a1d);	\draw[->] (a11d) -> (a1da);
		\draw[->] (a1d) -> (a0d);	\draw[->] (a1d) -> (a0da);
		\draw[->] (a2d) -> (a1da);	\draw[->] (a1da) -> (a0da);
		\node [blue] at ($(d4d.south) + (0,-0.1)$) {$P_{\text{VI}}$};
		\node [blue] at ($(a3d.south) + (-0.3,-0.1)$) {$P_{\text{V}}$};
		\node [blue] at ($(a11d.south) + (0,-0.1)$) {$P_{\text{III}}$};
		\node [blue] at ($(a1d.south) + (0,-0.1)$) {$P_{\text{III}}$};
		\node [blue] at ($(a0d.south) + (-0.2,-0.1)$) {$P_{\text{III}}$};
		\node [blue] at ($(a2d.south) + (0,-0.1)$) {$P_{\text{IV}}$};
		\node [blue] at ($(a1da.south) + (0,-0.1)$) {$P_{\text{II}}$};
		\node [blue] at ($(a0da.south) + (0,-0.1)$) {$P_{\text{I}}$};
	\end{tikzpicture}}
	\caption{Surface-type classification scheme for Painlev\'e equations}
\label{fig:Sakai-clsc-surf}
\end{figure}	
\end{center}

For any of the above surface types it is possible to construct a concrete geometric realization as follows. 
Start with a compact rational surface $\mathcal{X}$ obtained from blowing up
complex projective plane\footnote{This compactification of $\mathbb{C}^{2}$  is more convenient to study dynamics than the usual complex projective plane $\mathbb{P}_{\mathbb{C}}^{2}$; either one can be used except for the $\left(E_{8}^{(1)}\right)^{\text{c}}$ case that requires  $\mathbb{P}_{\mathbb{C}}^{2}$.}
$\mathbb{P}_{\mathbb{C}}^{1} \times \mathbb{P}_{\mathbb{C}}^{1}$
at \emph{eight} (possibly infinitely close) points $p_{1},\ldots p_{8}$ chosen in a special way. Namely 
the \emph{Picard lattice} and the anti-canonical class $-\mathcal{K}_{\mathcal{X}}$ of $\mathcal{X}$ are
\begin{equation*}
	\operatorname{Pic}(\mathcal{X}) = \operatorname{Span}_{\mathbb{Z}}\{\mathcal{H}_{1},\mathcal{H}_{2},
	\mathcal{E}_{1},\ldots \mathcal{E}_{8}\},\qquad 
	-\mathcal{K}_{\mathcal{X}} = 2\mathcal{H}_{1} + 2 \mathcal{H}_{2} - \mathcal{E}_{1} - \cdots - \mathcal{E}_{8}
\end{equation*}
where $\mathcal{H}_{i}$ are classes of the coordinate lines and $\mathcal{E}_{i}$ are classes of the exceptional divisors
$E_{i} = \pi^{-1}(p_{i})$ that are central fibers of the blowup map 
$\pi: \mathcal{X}\to \mathbb{P}_{\mathbb{C}}^{1} \times \mathbb{P}_{\mathbb{C}}^{1}$. There is always a 
bi-quadratic curve $\gamma\subset \mathbb{P}_{\mathbb{C}}^{1} \times \mathbb{P}_{\mathbb{C}}^{1}$, 
$[\gamma]\in 2\mathcal{H}_{1} + 2 \mathcal{H}_{2} = -\mathcal{K}_{\mathbb{P}_{\mathbb{C}}^{1} \times \mathbb{P}_{\mathbb{C}}^{1}}$,
possibly reducible, passing through any eight points, and we choose $\gamma$ and the points $p_{i}$ in such a way that 
the classes $\delta_{i}$ of the proper transforms $d_{i}$ of the components of $\gamma$ intersect according to the chosen
affine Dynkin diagram $\mathcal{R}$ and $\delta = -\mathcal{K}_{\mathcal{X}} = \sum_{i} m_{i} \delta_{i}$. In particular, this means that 
$\delta_{i}$ satisfy the \emph{root condition} $\delta_{i}^{2}=-2$ (note the opposite sign) and also $\delta^{2}=0$, i.e., 
we have an \emph{affine root system}. The 
multiplicities $m_{i}$ are also determined by $\mathcal{R}$ --- they are the components of the null-vector of its Cartan matrix.
Standard examples of such geometric realizations for each of the above $22$ cases, as well as careful explanations of the above 
constructions, are given in a recent excellent comprehensive survey \cite{KajNouYam:2017:GAPE}.

Some remarks are in order. First, points $p_{i}$ can be moved within their components of $\gamma$ without changing the surface type, 
so each configuration space is in fact a family $\{\mathcal{X}_{\mathbf{b}}\}_{\mathbf{b}\in \mathcal{B}}$, where parameters 
$\mathbf{b}$ are related to coordinates of the blowup points. Second, there are many possible different geometric realizations of
the same type, and in fact the first part of the identification question, after determining the surface type of the recurrence one
is interested in, which can be easily done by analyzing the singularities of the mapping defined by this recurrence, is to find an
isomorphism to one of the standard surfaces as listed in \cite{KajNouYam:2017:GAPE} or \cite{Sak:2001:RSAWARSGPE}. Often times 
at this point the recurrence is said to be equivalent to some standard discrete Painlev\'e equation having that 
surface type. However, there is no guarantee that this will be the case, since one needs to also consider the actual \emph{dynamics}. 

The origin of discrete Painlev\'e dynamics on the surface family $\{\mathcal{X}_{\mathbf{b}}\}$ is the extended affine Weyl group
of symmetries (Cremona isometries) of that family. Given the surface roots
$\Pi(\mathcal{R}) = \{\delta_{i}\}\subset \operatorname{Pic}(\mathcal{X})$, consider the corresponding root lattice 
$Q(\mathcal{R}) = \operatorname{Span}_{\mathbb{Z}}(\delta_{i})$ in $\operatorname{Pic}(\mathcal{X})$ 
and its orthogonal complement $Q(\mathcal{R})^{\perp}$. This orthogonal 
complement also has a \emph{root basis} $\Pi(\mathcal{R}^{\perp}) = \{\alpha_{i}\}$, $\alpha_{i}\bullet \delta_{j} = 0$ 
described by a \emph{dual} affine Dynkin diagram $\mathcal{R}^{\perp}$, whose type is 
known as the \emph{symmetry type} of the equation. Possible symmetry types are shown on Figure~\ref{fig:Sakai-clsc-symm} that is dual to Figure~\ref{fig:Sakai-clsc-surf}. The roots $\alpha_{i}$ are called the \emph{symmetry roots}, and they may have non-standard root lengths, as 
indicated on Figure~\ref{fig:Sakai-clsc-symm}, see \cite{Sak:2001:RSAWARSGPE} and \cite{KajNouYam:2017:GAPE} for details. 
The root lattice $Q(\mathcal{R}^{\perp}) = Q(\mathcal{R})^{\perp}$ is called the \emph{symmetry sub-lattice}. 
Since $(-\mathcal{K}_{\mathcal{X}})^{2} = 0$, we can restrict our attention to the sub-lattice 
$(-\mathcal{K}_{\mathcal{X}})^{\perp}\triangleleft \operatorname{Pic}(\mathcal{X})$, which is an affine root lattice of type $E_{8}^{(1)}$,
and both $Q(\mathcal{R}), Q(\mathcal{R}^{\perp})\triangleleft (-\mathcal{K}_{\mathcal{X}})^{\perp}$ with
$Q(\mathcal{R})\cap Q(\mathcal{R}^{\perp}) = \operatorname{Span}_{\mathbb{Z}}\{\delta = - \mathcal{K}_{\mathcal{X}}\}$.
\begin{center}
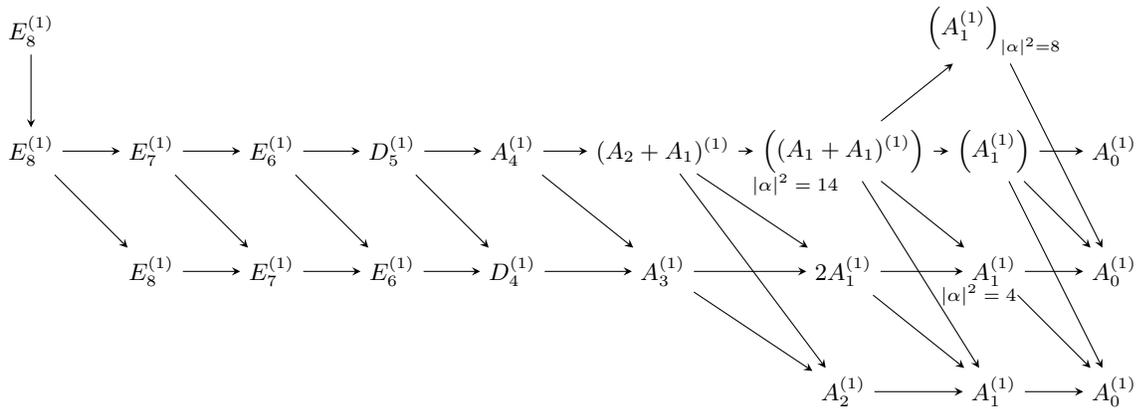
\begin{figure}[h]{\small
\begin{tikzpicture}[>=stealth,scale=0.8]
	\node (e8e) at (1,4) {$E_{8}^{(1)}$};
	\node (a1qa) at (17,4) {$\left(A_{1}^{(1)}\right)_{\scriptsize |\alpha|^{2}=8}$};
	\node (e8q) at (1,2) {$E_{8}^{(1)}$};
	\node (e7q) at (3,2) {$E_{7}^{(1)}$};
	\node (e6q) at (5,2) {$E_{6}^{(1)}$};
	\node (d5q) at (7,2) {$D_{5}^{(1)}$};
	\node (a4q) at (9,2) {$A_{4}^{(1)}$};
	\node (a21q) at (11.5,2) {$(A_{2} + A_{1})^{(1)}$};
	\node (a11q) at (14.5,2) {$\left((A_{1} + A_{1})^{(1)}\right)$};
	\node  at (13.7,1.45) {\scriptsize $|\alpha|^{2}=14$};
	\node (a1q) at (17,2) {$\left(A_{1}^{(1)}\right)$};
	\node (a0q) at (19,2) {$A_{0}^{(1)}$};
	\node (e8d) at (3,0) {$E_{8}^{(1)}$};
	\node (e7d) at (5,0) {$E_{7}^{(1)}$};
	\node (e6d) at (7,0) {$E_{6}^{(1)}$};
	\node (d4d) at (9,0) {$D_{4}^{(1)}$};
	\node (a3d) at (11.5,0) {$A_{3}^{(1)}$};
	\node (a11d) at (14.5,0) {$2A_{1}^{(1)}$};
	\node (a1d) at (17,0) {$A_{1}^{(1)}$};
	\node  at (16.75,-0.4) {\scriptsize $|\alpha|^{2}=4$};	
	\node (a0d) at (19,0) {$A_{0}^{(1)}$};
	\node (a2d) at (14.5,-2) {$A_{2}^{(1)}$};
	\node (a1da) at (17,-2) {$A_{1}^{(1)}$};
	\node (a0da) at (19,-2) {$A_{0}^{(1)}$};
	\draw[->] (e8e) -> (e8q);	\draw[->] (a1qa) -> (a0d); 	
	\draw[->] (e8q) -> (e7q); 	\draw[->] (e8q) -> (e8d); 	
	\draw[->] (e7q) -> (e6q); 	\draw[->] (e7q) -> (e7d); 	
	\draw[->] (e6q) -> (d5q); 	\draw[->] (e6q) -> (e6d); 	
	\draw[->] (d5q) -> (a4q); 	\draw[->] (d5q) -> (d4d); 	
	\draw[->] (a4q) -> (a21q); 	\draw[->] (a4q) -> (a3d); 	
	\draw[->] (a21q) -> (a11q); \draw[->] (a21q) -> (a11d); \draw[->] (a21q) -> (a2d);	
	\draw[->] (a11q) -> (a1q); 	\draw[->] (a11q) -> (a1d); 	\draw[->] (a11q) -> (a1qa); 	\draw[->] (a11q) -> (a1da);
	\draw[->] (a1q) -> (a0q); 	\draw[->] (a1q) -> (a0d);	\draw[->] (a1q) -> (a0da);
	\draw[->] (e8d) -> (e7d);
	\draw[->] (e7d) -> (e6d);
	\draw[->] (e6d) -> (d4d);
	\draw[->] (d4d) -> (a3d);	
	\draw[->] (a3d) -> (a11d);	\draw[->] (a3d) -> (a2d);	
	\draw[->] (a11d) -> (a1d);	\draw[->] (a11d) -> (a1da);	
	\draw[->] (a1d) -> (a0d);	\draw[->] (a1d) -> (a0da);	
	\draw[->] (a2d) -> (a1da);	\draw[->] (a1da) -> (a0da);	
	\end{tikzpicture}}
	\caption{Symmetry-type classification schemes for Painlev\'e equations}
\label{fig:Sakai-clsc-symm}
\end{figure}
\end{center}

Given an affine root system, in our case $\Pi(\mathcal{R}^{\perp})$, it is possible to consider the corresponding affine 
Weyl group $W(\mathcal{R}^{\perp})$ whose generators $w_{\alpha}$ are the nodes of the Dynkin diagram $\mathcal{R}^{\perp}$ 
and relations correspond to the edges. E.g., in the simply-laced case, generators corresponding to connected nodes 
$\alpha_{i}$, $\alpha_{j}$ should satisfy the \emph{braid relation} 
$w_{\alpha_{i}}w_{\alpha_{j}}w_{\alpha_{i}} = w_{\alpha_{j}}w_{\alpha_{i}}w_{\alpha_{j}}$. In the Painlev\'e case, 
there is a representation of $W(\mathcal{R}^{\perp})$ on $\operatorname{Pic}(\mathcal{X})\otimes_{\mathbb{Z}} \mathbb{Q}$ 
given by reflection actions in the symmetry roots $\alpha$,
\begin{equation*}
w_{\alpha}(\mathcal{C}) = \mathcal{C} - 2 \frac{\mathcal{C}\bullet \alpha}{\alpha\bullet \alpha}\alpha,\qquad 
\mathcal{C}\in \operatorname{Pic}(\mathcal{X}), \quad \alpha\in \Pi(\mathcal{R}^{\perp}).	
\end{equation*} 
This group and its representation can be extended by the group of automorphisms of the Dynkin diagram $\mathcal{R}^{\perp}$
(or, equivalently, of $\mathcal{R}$) to give the \emph{extended Weyl group} 
$\widetilde{W}(\mathcal{R}^{\perp}) = W(\mathcal{R}^{\perp})\rtimes \operatorname{Aut}(\mathcal{R}^{\perp})$. The action 
of $\widetilde{W}(\mathcal{R}^{\perp})$ \emph{can be extended} from $\operatorname{Pic}(\mathcal{X})$ to the family of 
surfaces $\{\mathcal{X}_{\mathbf{b}}\}$ --- for each $w_{\alpha}$ one can construct the birational map 
$\varphi_{\alpha}:\mathbb{P}_{\mathbb{C}}^{1} \times \mathbb{P}_{\mathbb{C}}^{1}\dashrightarrow 
\mathbb{P}_{\mathbb{C}}^{1} \times \mathbb{P}_{\mathbb{C}}^{1}$ that lifts to an isomorphism 
$\varphi_{\alpha}: \mathcal{X}_{b}\to \mathcal{X}_{\varphi(b)}$ between two surfaces in the family so that the induced map 
$(\varphi_{\alpha})^{*} = w_{\alpha}$. Similar extensions can be constructed for the automorphisms as well. This isomorphism
is generically non-autonomous, i.e., parameters of the family change and we indeed get some dynamics on the family. 
This representation is known as \emph{birational representation} of 
$\widetilde{W}(\mathcal{R}^{\perp})$.

We are finally in the position to give the \emph{geometric definition} of a \emph{discrete Painlev\'e equation}. It is 
a discrete dynamical system on the family of surfaces $\{\mathcal{X}_{\mathbf{b}}\}$ that corresponds to a 
\emph{translational element} $\mathbf{t}\in \widetilde{W}(\mathcal{R}^{\perp})$. 
The resulting action gives a translation (by the multiple of the 
anti-canonical class $\delta = [-K_{\mathcal{X}}]$) 
in the symmetry sub-lattice, so we can encode it by the translation action on the symmetry roots 
$\Pi(\mathcal{R}^{\perp})$. Further, there exist canonical parameters, known as the \emph{root variables},
that are compatible with the group action, and so we can also encode the translation element completely 
by the changes in the root variables. Finally, any translation element can be written as a word in the 
generators of $\widetilde{W}(\mathcal{R}^{\perp})$. However, conjugating the translation by some 
element of $\widetilde{W}(\mathcal{R}^{\perp})$ gives an equivalent dynamics, so we really should consider
its \emph{equivalence class}. So abstractly a discrete Painlev\'e equation then is a triple 
$(\mathcal{R}, \mathcal{R}^{\perp}, [\mathbf{t}])$ that we should also consider with some standard 
geometric realization, as in \cite{KajNouYam:2017:GAPE}, on the family of Sakai surfaces $\{\mathcal{X}_{\mathbf{b}}\}$. 

The family $\mathcal{X}_{\mathbf{b}}$ is then partitioned into the equivalence classes of orbits w.r.t.~this dynamic $[\mathbf{t}]$.
It turns out that some orbits can correspond to \emph{special point configurations} within the same class, and the
resulting equations would have properties that are different from generic cases. One extreme example are \emph{stationary} orbits
or \emph{autonomous} versions of the discrete Painlev\'e equations known as the QRT maps \cite{CarDzhTak:2017:FDIMDPE}. 
The other example is the familiar case in the differential Painlev\'e theory when equations admit algebraic solutions or 
solutions that can be expressed in terms of special functions for \emph{special values of parameters}, the same is true for 
discrete Painlev\'e equations as well.
Finally, special parameter values can result in the appearance of nodal curves, which can change
the symmetry group.
Thus, only the surface type classification 
of Painlev\'e equations shown on Figure~\ref{fig:Sakai-clsc-surf} is complete, and the symmetry type classification is 
only \emph{generic} and does not include possible special cases. Closely related to that is the notion of 
\emph{projective reductions} --- these are dynamical systems that are defined by elements of infinite order that are not
translations on the symmetry lattice, but whose powers are. These systems can again become 
translations on the symmetry lattice for a special symmetry group for special parameter values, 
\cite{KajNakTsu:2011:PRDPSTA}, \cite{AtkHowJosNak:2016:GEDER}
\cite{CarDzhTak:2017:FDIMDPE}.

Thus, the full version of the \emph{Painlev\'e identification problem} for a nonlinear second order recurrence 
appearing in an applied problem, as outlined in \cite{DzhFilSto:2020:RCDOPWHWDPE}, is the following. 
\begin{itemize}
	\item Use the singularity structure of the recurrence to determine whether it 
	is of discrete Painlev\'e class and if so, determine its surface type.
	\item Find an isomorphism from the resulting surface to the standard realization of the family, as in 
	\cite{KajNouYam:2017:GAPE}.
	\item Using that and the action of the dynamic on the Picard lattice, 
	find the equivalence class of the translation element, or translation vector.
	\item See if the parameters of the applied problem in fact result in a special configuration of points that 
	can change the symmetry group of the dynamic, and if so, find that symmetry group.
\end{itemize}
With such detailed information we then will be in the position to compare the recurrence relation not only to the standard
discrete Painlev\'e equations, but also to other examples, e.g., corresponding to different weights, that have the same 
surface type.


\section{Gaussian Weight with a Single Jump} 
\label{sec:GW-jump}

We now apply the above scheme to the recurrence relation obtained by C.~Min and Y.~Chen in \cite{MinChe:2019:PTHDGDGW}
in studying polynomials $\{P_{n}\}$ that are orthogonal w.r.t.~the Gaussian weight with a single jump:
\begin{equation}\label{eq:GW-jump}
    w(x, t) = e^{-\nu(x)} (A + B\, \theta(x - t)), \quad \text{where } A \geq 0,\  A + B \geq 0,\ \nu(x) = x^{2},
\end{equation}
and $\theta(x)$ is the usual Heaviside step function; i.e., $\theta(x) = 1$ for $x>0$ and $0$ otherwise.
This weight has also been studied previously by Y.~Chen and G.~Pruessner for $A=1 - {\beta}/{2}$, $B=\beta$ \cite{ChePru:2005:OPWDW} and 
by A.~Its for $A = e^{\mathfrak{i} \beta \pi}$, $B = e^{- \mathfrak{i} \beta \pi} - e^{\mathfrak{i} \beta \pi}$ \cite{Its:2011:DPEOP}.

So consider a family of \emph{monic} polynomials that are orthogonal with respect to the weight \eqref{eq:GW-jump}
\begin{equation}
    \int^{\infty}_{-\infty} P_k(x,t) P_l(x,t) w(x,t) dx = h_k(t) \delta_{kl}, \qquad k,l\in \mathbb{Z}_{\geq 0}.
\end{equation}
Here $h_{k}(t) = \|P_{k}(x,t)\|_{L^{2}({\omega}(x,t)\,dx)}^{2}$.
While studying such orthogonal polynomial ensembles, discrete Painlev\'e equations usually describe the dynamics w.r.t.~the 
parameter $n$ of either the coefficients $\alpha_{n}$, $\beta_{n}$ of the three-term recurrence relation
\begin{equation*}
    x P_n(x,t) = P_{n+1}(x,t) + \alpha_{n} (t) P_n(x,t) + \beta_{n} (t) P_{n-1}(x,t),
\end{equation*}
or, as in this case, the (related) coefficients of the ladder operators $A_n$, $B_n$:
\begin{equation*}
    \begin{aligned}
            \frac{d}{dz} P_n(z,t) &= \beta_n(t) A_n(z,t) P_{n-1} (z,t) - B_n(z,t) P_n(z,t), \\
            \frac{d}{dz} P_{n-1}(z,t) &= (B_n(z,t) + \nu'(z)) P_{n-1} (z,t) - A_{n-1}(z,t) P_n(z,t).
    \end{aligned}
\end{equation*}
Chen and Min showed that these operators can be parameterized as 
\begin{equation*}
    A_n(z) = 2 + \frac{R_n(t)}{z-t}, \quad B_n(z) = \frac{r_n(t)}{z-t},
\end{equation*}
where
\begin{equation*}
    R_n(t) = \frac{B P_n^2(t,t) e^{-t^2}}{h_n(t)}, \quad r_n(t) = \frac{B P_n(t,t) P_{n-1}(t,t) e^{-t^2}}{h_{n-1}(t)},
\end{equation*}
and they also found recurrence relations satisfied by the coefficients $r_{n}(t)$ and $R_{n}(t)$. 
We change the variable $t$ in this mapping to the variable $s$ since we at this point can not be sure whether 
it coincides with the parameter $t$ that often appears in discrete Painlev\'e equations in relations to continuous Painlev\'e 
equations (in fact, we will see that it needs rescaling), and also for convenience we change the notation,
$x_n(s) = r_n(s)$, $y_n(s) = R_n(s)$. Then the recurrences \cite[(2.20-2.21)]{MinChe:2019:PTHDGDGW} take the form
\begin{equation}\label{eq:MinChen}
		x_{n+1} + x_n =  \left(s - \frac{y_n}{2} \right) y_{n}, \quad
		x_{n}^2 = \left(\frac{n + x_n}{2} \right) y_n  y_{n-1}.
\end{equation}
This is the dynamics that we want to study. Our main result is the following Theorem.

\begin{theorem}\label{eq:math-thm}
	The change of variables 
	\begin{equation}\label{eq:var-change}
		\left\{
		\begin{aligned}
			q(x,y;s)&= - \frac{\sqrt{2}x}{y},\\
			p(x,y;s)&=- \frac{2x + y(y-2s)}{\sqrt{2} y},\\
			t(s) &= \sqrt{2}s;
		\end{aligned}
		\right. \qquad
		\left\{
		\begin{aligned}
			x(q,p;t)&=q(p-q-t),\\
			y(q,p,t)&=\sqrt{2}(q-p+t),\\
			s(t)&=\frac{t}{\sqrt{2}},
		\end{aligned}
		\right. \qquad
	\end{equation}
	identifies the recurrence relation \eqref{eq:MinChen} with the standard discrete Painlev\'e equation 
	\begin{equation}\label{eq:dP-std}
			\overline{q} + q = p - t - \frac{a_2}{p},\quad
			p + \underline{p} = q + t + \frac{a_1}{q},\qquad \overline{a}_{1}=a_{1}-1,\ \overline{a}_{2}=a_{2}+1.		
	\end{equation}
	on the $E_{6}^{(1)}$ Sakai surface with the constrained point configuration corresponding to the vanishing of 
	the root variable $a_{0}$; $a_{1}+a_{2}=1$, $(a_{0}=0)$.
	Thus, the orbit corresponding to this recurrence in the full parameter space of Sakai surfaces of type $E_{6}^{(1)}$
	has to be contained in the surface family over the restricted parameter space satisfying $a_{0}=0$ (i.e., the nodal
	curve condition has to be preserved). As a result, the symmetry group $\mathcal{G}$ of this constrained family is smaller than 
	the full group $\widetilde{W}\left(A_{2}^{(1)}\right)$; in fact, $\mathcal{G}\simeq W\left(A_{1}^{(1)}\right)$.
\end{theorem}
This Theorem will be proved in Section~\ref{sub:match}, with the exception of the computation fo the symmetry group which is 
beyond the scope of this note.


\section{The Identification Procedure} 
\label{sec:identification}
In this section we regularize the dynamics given by equations \eqref{eq:MinChen} using the blowups and then follow the steps described in 
\cite{DzhFilSto:2020:RCDOPWHWDPE} to match it with one of the standard discrete Painlev\'e equations. We begin by 
constructing the configuration space, or the family of Sakai surfaces, for this mapping. 

\subsection{The configuration space} 
\label{sub:conf-space}
Let us now understand the singularity structure of the mapping \eqref{eq:MinChen}. Since this process closely follows the 
procedure described in detail in \cite{DzhFilSto:2020:RCDOPWHWDPE}, we omit most of the computations and just state the results. 

As it is often the case, the recurrence relations \eqref{eq:MinChen} naturally defines two mappings: $\psi_1^{(n)}: (x_n, y_n) \to (x_{n+1}, y_n)$ 
and $\psi_2^{(n)}: (x_n, y_n) \to (x_n, y_{n-1})$ obtained from the first and the second equations respectively. 
The full \emph{forward step} of the dynamic is given by the composed mapping 
$\psi_f^{(n)} = (\psi_2^{(n+1)})^{-1} \circ \psi_1^{(n)}: (x_n,y_n) \to (x_{n+1},y_{n+1})$ and the 
\emph{backward step} is $\psi_b^{(n)} = (\psi_1^{(n-1)})^{-1} \circ \psi_2^{(n)}: (x_n,y_n) \to (x_{n-1},y_{n-1})$. 
Using the standard notation 
$x:=x_{n}$, $\overline{x}:=x_{n+1}$, $\underline{x}:=x_{n-1}$, and similarly for $y$, as well as abbreviating 
$\psi=\psi_{f}^{(n)}$, $\psi^{-1} = \psi_{b}^{(n)}$, 
we see that the mapping that we are interested in takes the form
\begin{equation}\label{eq:MC-forward}
	\psi:\quad\left\{ \begin{aligned}
		\overline{x} &=s y - \frac{y^{2}}{2}- x, \\
		\overline{y} &= \frac{(y^2 - 2 s y + 2 x)^2}{y  (2 (n + 1 -x) -y(y-2s))},
	\end{aligned}\right.
\end{equation}
We compute the basepoints of this mapping to be 
\begin{align*}
		q_{1}\left(x = 0, y = 0\right) &\leftarrow  q_{2}\left(u_{1} = x = 0, v_{1}= \frac{y}{x} = 0\right),\quad  q_{3}\left(x = -n, Y = \frac{1}{y} = 0\right),
		\\
		q_{4}\left(X = \frac{1}{x} = 0, Y = 0 \right) &\leftarrow q_{5}\left(U_{4} = \frac{X}{Y} = 0, V_{4} = Y = 0\right)\\
		&\leftarrow
		q_{6}\left(u_{5} = U_4 = 0, v_{5} = \frac{V_4}{U_4} = - \frac{1}{2} \right)\\
		&\leftarrow
		q_{7}\left( u_6 = u_5 = 0, v_6 = \frac{1 + 2 v_5}{2 u_5} = - \frac{s}{2} \right) \\
		&\leftarrow
		q_{8}\left(u_{7} = u_6 = 0, v_{7} = \frac{2 v_6 + s}{2 u_6} = \frac{1}{4} (n + 1 - 2 s^2) \right).
\end{align*}

\begin{center}
\begin{figure}[ht]
        \centering
	\begin{tikzpicture}[scale=0.9, >=stealth,basept/.style={circle, draw=red!100, fill=red!100, thick, inner sep=0pt,minimum size=1.2mm}, scale=1.3]
	\begin{scope}[xshift=0cm,yshift=0cm]
	\draw [black, line width = 1pt] (-0.4,0) -- (2.4,0)	node [pos=0,left] {\small $H_{y}$} node [pos=1,right] {\small $y=0$};
	\draw [black, line width = 1pt] (-0.4,2.5) -- (2.4,2.5) node [pos=0,left] {\small $H_{y}$} node [pos=1,right] {\small $y=\infty$};
	\draw [black, line width = 1pt] (0,-0.4) -- (0,2.9) node [pos=0,below] {\small $H_{x}$} node [pos=1,above] {\small $x=0$};
	\draw [black, line width = 1pt] (2.0,-0.4) -- (2.0,2.9) node [pos=0,below] {\small $H_{x}$} node [pos=1,above] {\small $x=\infty$};
	\node (q1) at (0,0) [basept,label={[xshift = -8pt, yshift=-15pt] \small $q_{1}$}] {};
	\node (q2) at (0.7,-0.3) [basept,label={[yshift=-15pt] \small $q_{2}$}] {};
	\node (q3) at (1,2.5) [basept,label={[yshift=-15pt] \small $q_{3}$}] {};
	\node (q4) at (2,2.5) [basept,label={[xshift = -8pt, yshift=-15pt] \small $q_{4}$}] {};
	\node (q5) at (2.5,1.9) [basept,label={[xshift=10pt,yshift=-8pt] \small $q_{5}$}] {};
	\node (q6) at (2.5,1.4) [basept,label={[xshift=10pt,yshift=-8pt] \small $q_{6}$}] {};
	\node (q7) at (2.5,0.9) [basept,label={[xshift=10pt,yshift=-8pt] \small $q_{7}$}] {};
	\node (q8) at (2.5,0.4) [basept,label={[xshift=10pt,yshift=-8pt] \small $q_{8}$}] {};
	\draw [red, line width = 0.8pt, ->] (q2) -- (0.5,0) -- (q1);
	\draw [red, line width = 0.8pt, ->] (q5) -- (2,2.1) -- (q4);
	\draw [red, line width = 0.8pt, ->] (q6) -- (q5);
	\draw [red, line width = 0.8pt, ->] (q7) -- (q6);
	\draw [red, line width = 0.8pt, ->] (q8) -- (q7);
	\end{scope}
	\draw [->] (5.5,1.5)--(3.5,1.5) node[pos=0.5, below] {$\operatorname{Bl}_{q_{1}\cdots q_{8}}$};
	\begin{scope}[xshift=6.3cm,yshift=0cm]
	\draw [blue, line width = 1pt] (0.4,0) -- (3.2,0) node [pos=1, right] {\small $H_{y}-F_{1}-F_{2}$};
	\draw [blue, line width = 1pt] (-0.4,2.5) -- (4,2.5) node [pos=0, left] {\small $H_{y}-F_{3}-F_{4}$};
	\draw [red, line width = 1pt] (0,-0.4) -- (0,2.9) node [pos=1, above] {\small $H_{x}-F_{1}$};
	\draw [blue, line width = 1pt] (-0.3,0.3) -- (0.7,-0.7) node [pos=0, left] {\small $F_{1}-F_{2}$};
	\draw [red, line width = 1pt] (0.3,-0.7) -- (1.2,0.2) node [pos=0, below] {\small $F_{2}$};		
	\draw [red, line width = 1pt] (1.5,-0.4) -- (1.5,2.1) node [pos=0, xshift=0pt,yshift=-5pt] {\small $H_{x}-F_{3}$};
	\draw [red, line width = 1pt] (1.3,1.7) -- (2.3,2.7) node [pos=0, left] {\small $F_{3}$};	
	\draw [blue, line width = 1pt] (3,-0.7) -- (3,1.7) node [pos=0, below] {\small $H_{x}-F_{4}-F_{5}$};
	\draw [blue, line width = 1pt] (3,2.9) -- (4,1.9) node [pos=0, above] {\small $F_{4}-F_{5}$};
	\draw [blue, line width = 1pt] (3.9,2.3) -- (2.8,1.2) node [pos=1, xshift=-15pt,yshift=-5pt] {\small $F_{5}-F_{6}$};
	\draw [blue, line width = 1pt] (3,2) -- (3.8,1.2) node [pos=0, xshift=-15pt,yshift=2pt] {\small $F_{6}-F_{7}$};
	\draw [blue, line width = 1pt] (4.2,1.9) -- (3.5,1.2) node [pos=0, right] {\small $F_{7}-F_{8}$};
	\draw [red, line width = 1pt] (3.7,1.8) -- (4.3,1.2) node [pos=1, below] {\small $F_{8}$};
	\end{scope}
	\end{tikzpicture}
	\caption{The Sakai surface for the Chen-Min recurrence relation}
	\label{fig:MC-surf}
\end{figure}
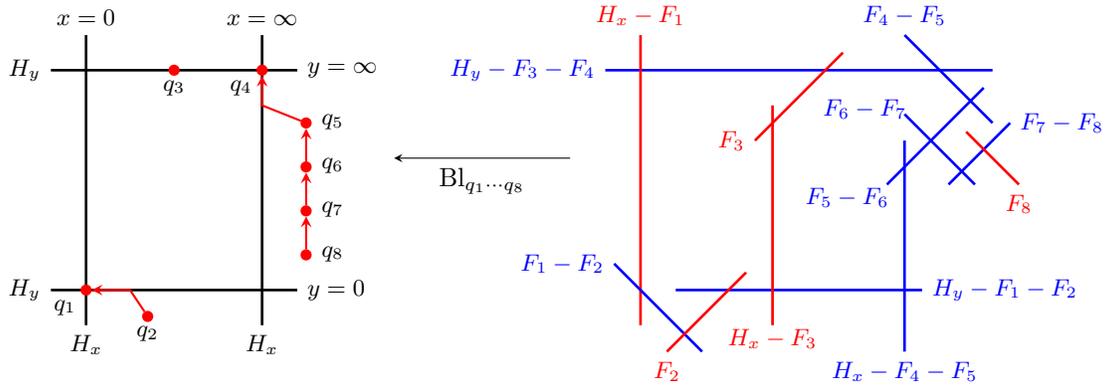
\end{center}

\begin{figure}[ht]
    \begin{equation}\label{eq:MC-surf-roots}			
	    \raisebox{-45.0pt}{\begin{tikzpicture}[scale=1.5, elt/.style={circle, draw=black!100, thick, inner sep=0pt, minimum size=2mm}]
            \path (-1.2,-0.6) node (d1) [elt] {} 
            (-0.6,-0.3) node (d2) [elt] {}
            (0,0) node (d3) [elt] {}
            ( 0.6,-0.3) node (d4) [elt] {}
            ( 1.2,-0.6) node (d5) [elt] {}
            ( 0,0.67) node (d6) [elt] {}
            ( 0,1.34) node (d0) [elt] {};
            \draw [black] (d1) -- (d2) -- (d3) -- (d4) -- (d5) (d3) -- (d6) -- (d0);
            \node at ($(d1.west) + (-0.2,0.15)$) {$\delta_{1}$};
            \node at ($(d2.west) + (-0.2,0.15)$) {$\delta_{2}$};
            \node at ($(d3.east) + (0.2,0.15)$) {$\delta_{3}$};
            \node at ($(d4.east) + (0.2,0.15)$) {$\delta_{4}$};
            \node at ($(d5.east) + (0.2,0.15)$) {$\delta_{5}$};
            \node at ($(d6.east) + (0.2,0)$) {$\delta_{6}$};
            \node at ($(d0.east) + (0.2,0)$) {$\delta_{0}$};
	    \end{tikzpicture}} \qquad
		\begin{alignedat}{2}
			\delta_{0} &= \mathcal{H}_{y} - \mathcal{F}_{1} - \mathcal{F}_{2} , &\qquad  \delta_{4} &= \mathcal{F}_{4} - \mathcal{F}_{5}, \\
			\delta_{1} &= \mathcal{F}_{7} - \mathcal{F}_{8}, &\qquad  \delta_{5} &= \mathcal{H}_{y} - \mathcal{F}_{3} - \mathcal{F}_{4}, \\
			\delta_{2} &= \mathcal{F}_{6} - \mathcal{F}_{7} , &\qquad  \delta_{6} &= \mathcal{H}_{x} - \mathcal{F}_{4} - \mathcal{F}_{5}.\\
			\delta_{3} &= \mathcal{F}_{5} - \mathcal{F}_{6},
		\end{alignedat}
    \end{equation}
	\caption{The surface root basis for the Chen-Min recurrence} \label{fig:MC-surf-roots}	
\end{figure}
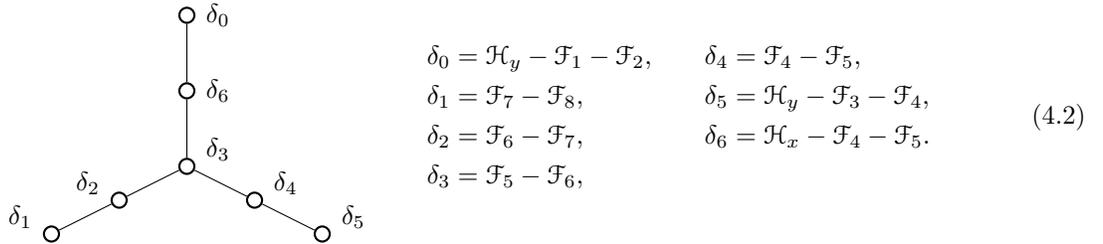

Resolving these basepoints using the blowup procedure, we get the Sakai surface $\mathcal{X}$, or the configuration space of this recurrence, shown on Figure~\ref{fig:MC-surf}.
Looking at the configuration of the $-2$-curves that are irreducible components of the anti-canonical divisor, we immediately see that the 
surface type of this recurrence is $E_{6}^{(1)}$, with the surface 
root basis\footnote{There is an obvious ambiguity in the labeling of the surface roots, here it is chosen to give the correct 
identification with the standard case, see \cite{DzhFilSto:2020:RCDOPWHWDPE} for how to adjust it if necessary.} 
shown on Figure~\ref{fig:MC-surf-roots}.

We also notice that there is another $-2$-curve $F_{1} - F_{2}$ that is \emph{disjoint} from the anti-canonical divisor. Such curves 
are called \emph{nodal curves} and their appearance indicates that we have a special point configuration whose properties,
including the symmetry group, can be different from the generic case. 

Finally, we can compute the evolution of the surface roots \eqref{eq:MC-surf-roots} under the mapping 
$\psi_{*}:\operatorname{Pic}(\mathcal{X})\to \operatorname{Pic}(\mathcal{X})$ induced by \eqref{eq:MC-forward} to be
\begin{equation}\label{eq:MinChen-evol}
	\begin{aligned}
	    \mathcal{H}_{x}& \mapsto 4 \mathcal{H}_{x} + 2 \mathcal{H}_{y} - 2 \mathcal{F}_{123} - \mathcal{F}_{4567}, \qquad& 
		\mathcal{F}_{4} &\mapsto 2\mathcal{H}_{x} + \mathcal{H}_{y} - \mathcal{F}_{12347},  \\
	    \mathcal{H}_{y}& \mapsto 2\mathcal{H}_{x} + \mathcal{H}_{y} - \mathcal{F}_{1234}, \qquad&  
		\mathcal{F}_{5} &\mapsto 2\mathcal{H}_{x} + \mathcal{H}_{y} - \mathcal{F}_{12346}, \\
	    \mathcal{F}_{1} &\mapsto \mathcal{H}_{x}  - \mathcal{F}_{2}, \qquad& 
		\mathcal{F}_{6} &\mapsto 2\mathcal{H}_{x} + \mathcal{H}_{y} - \mathcal{F}_{12345}, \\
	    \mathcal{F}_{2} &\mapsto \mathcal{H}_{x} - \mathcal{F}_{1}, \qquad& 
		\mathcal{F}_{7} &\mapsto \mathcal{H}_{x} + \mathcal{H}_{y} - \mathcal{F}_{123}, \\
		\mathcal{F}_{3} &\mapsto \mathcal{F}_{8}, \qquad& 	
		\mathcal{F}_{8} &\mapsto \mathcal{H}_{x} - \mathcal{F}_{3},		
	\end{aligned}
\end{equation}
where we use the notation $\mathcal{F}_{i\cdots j} = \mathcal{F}_{i} + \cdots + \mathcal{F}_{j}$.			
This is the dynamic that we now want to match with the standard case. First let us recall the standard geometric realization 
of the $E_{6}^{(1)}$ Sakai surface following \cite{KajNouYam:2017:GAPE}.


\subsection{The standard geometric realization of the $E_{6}^{(1)}$ Sakai surface} 
\label{sub:E6-standard}

Consider the point configuration and the resulting Sakai surface shown on Figure~\ref{fig:std-surf}.
\begin{center}
\begin{figure}[ht]
        \centering
	\begin{tikzpicture}[scale=0.9, >=stealth,basept/.style={circle, draw=red!100, fill=red!100, thick, inner sep=0pt,minimum size=1.2mm}, scale=1.3]
	\begin{scope}[xshift=0cm,yshift=0cm]
	\draw [black, line width = 1pt] (-0.4,0) -- (2.9,0)	node [pos=0,left] {\small $H_{p}$} node [pos=1,right] {\small $p=0$};
	\draw [black, line width = 1pt] (-0.4,2.5) -- (2.9,2.5) node [pos=0,left] {\small $H_{p}$} node [pos=1,right] {\small $p=\infty$};
	\draw [black, line width = 1pt] (0,-0.4) -- (0,2.9) node [pos=0,below] {\small $H_{q}$} node [pos=1,above] {\small $q=0$};
	\draw [black, line width = 1pt] (2.5,-0.4) -- (2.5,2.9) node [pos=0,below] {\small $H_{q}$} node [pos=1,above] {\small $q=\infty$};
	\node (p1) at (2.5,0) [basept,label={[xshift = 8pt, yshift=-15pt] \small $p_{1}$}] {};
	\node (p2) at (2,0.5) [basept,label={[yshift=0pt] \small $p_{2}$}] {};
	\node (p3) at (0,2.5) [basept,label={[xshift = -8pt, yshift=-15pt] \small $p_{3}$}] {};
	\node (p4) at (0.5,2) [basept,label={[xshift = 0pt, yshift=-15pt] \small $p_{4}$}] {};
	\node (p5) at (2.5,2.5) [basept,label={[xshift=-8pt,yshift=-15pt] \small $p_{5}$}] {};
	\node (p6) at (3,2) [basept,label={[xshift=10pt,yshift=-8pt] \small $p_{6}$}] {};
	\node (p7) at (3,1.3) [basept,label={[xshift=10pt,yshift=-8pt] \small $p_{7}$}] {};
	\node (p8) at (3,0.6) [basept,label={[xshift=10pt,yshift=-8pt] \small $p_{8}$}] {};
	\draw [red, line width = 0.8pt, ->] (p2) -- (p1);
	\draw [red, line width = 0.8pt, ->] (p4) -- (p3);
	\draw [red, line width = 0.8pt, ->] (p6) -- (p5);
	\draw [red, line width = 0.8pt, ->] (p7) -- (p6);
	\draw [red, line width = 0.8pt, ->] (p8) -- (p7);
	\end{scope}
	\draw [->] (6,1)--(4,1) node[pos=0.5, below] {$\operatorname{Bl}_{p_{1}\cdots p_{8}}$};
	\begin{scope}[xshift=6.3cm,yshift=0cm]
	\draw [red, line width = 1pt] (-0.4,0) -- (2.6,0) node [pos=0, left] {\small $H_{p}-E_{1}$};
	\draw [blue, line width = 1pt] (0.4,2.5) -- (2.7,2.5) node [pos=0, left] {\small $H_{p}-E_{3}-E_{5}$};
	\draw [red, line width = 1pt] (0,-0.4) -- (0,2.1) node [pos=0, below] {\small $H_{q}-E_{2}$};
	\draw [blue, line width = 1pt] (3,0.4) -- (3,2.1) node [pos=0.5, right] {\small $H_{q}-E_{1}-E_{5}$};
	\draw [blue, line width = 1pt] (-0.2,1.7) -- (0.8,2.7) node [pos=0, left] {\small $E_{3}-E_{4}$};
	\draw [red, line width = 1pt] (0.2,2.3) -- (0.9,1.6) node [pos=1, below] {\small $E_{4}$};	
	\draw [blue, line width = 1pt] (2.2,-0.2) -- (3.2,0.8) node [pos=0, below] {\small $E_{1}-E_{2}$};	
	\draw [red, line width = 1pt] (2.8,0.2) -- (2.1,0.9) node [pos=1, above] {\small $E_{2}$};
	\draw [blue, line width = 1pt] (3.2,1.7) -- (2.2,2.7) node [pos=1, xshift=-10pt, yshift=5pt] {\small $E_{5}-E_{6}$};
	\draw [blue, line width = 1pt] (2.5,2) -- (3.5,3) node [pos=0, left] {\small $E_{6}-E_{7}$};	
	\draw [blue, line width = 1pt] (3.9,2) -- (2.9,3) node [pos=0, xshift=10pt, yshift=-5pt] {\small $E_{7}-E_{8}$};
	\draw [red, line width = 1pt] (3.3,2.1) -- (4.0,2.8) node [pos=1, right] {\small $E_{8}$};	
	\end{scope}
	\end{tikzpicture}
	\caption{The standard $E_{6}^{(1)}$ Sakai surface}
	\label{fig:std-surf}
\end{figure}
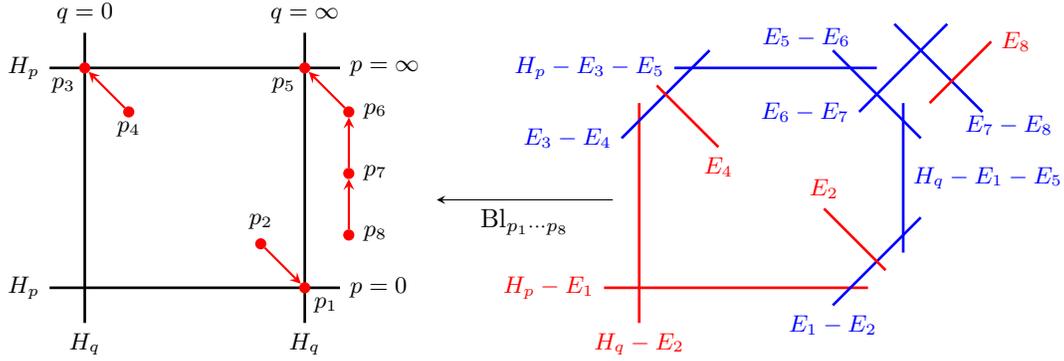
where the coordinates of the basepoints are given in terms of the \emph{root variables} $a_{0}$, $a_{1}$, and 
$a_{2}$ constrained by the normalization condition $a_{0}+a_{1}+a_{2}=1$ as follows:
\begin{equation}\label{eq:std-basepts}
	\begin{aligned}
		p_{1}\left(Q = \frac{1}{q}=0, p = 0\right) &\leftarrow  p_{2}\left(u_{1} = Q = 0, v_{1}= \frac{p}{Q} = -a_{2}\right),\\  
		p_{3}\left(q = 0, P = \frac{1}{p} = 0\right) &\leftarrow  p_{4}\left(U_{3} = \frac{q}{P} = a_{1}, V_{1}= P = 0\right)\\
		p_{5}\left(Q =  0, P = 0 \right) &\leftarrow p_{6}\left(u_{5} = Q = 0, v_{5} = \frac{P}{Q} = 1\right)\\
		&\leftarrow  
		p_{7}\left( u_{6} = u_{5} = 0, v_{6} = \frac{v_{5}-1}{u_{5}} = - t \right) \\
		&\leftarrow 
		p_{8}\left(u_{7} = u_{6} = 0, v_{7} = \frac{v_6 + t}{u_6} = a_{0} + t^{2} \right).
	\end{aligned}
\end{equation}
\end{center}

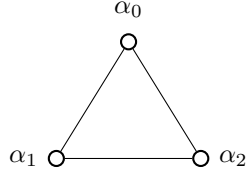
\begin{figure}[ht]
    \begin{equation}\label{eq:std-surf-roots}			
	    \raisebox{-45.0pt}{\begin{tikzpicture}[scale=1.5, elt/.style={circle, draw=black!100, thick, inner sep=0pt, minimum size=2mm}]
            \path (-1.2,-0.6) node (d1) [elt] {} 
            (-0.6,-0.3) node (d2) [elt] {}
            (0,0) node (d3) [elt] {}
            ( 0.6,-0.3) node (d4) [elt] {}
            ( 1.2,-0.6) node (d5) [elt] {}
            ( 0,0.67) node (d6) [elt] {}
            ( 0,1.34) node (d0) [elt] {};
            \draw [black] (d1) -- (d2) -- (d3) -- (d4) -- (d5) (d3) -- (d6) -- (d0);
            \node at ($(d1.west) + (-0.2,0.15)$) {$\delta_{1}$};
            \node at ($(d2.west) + (-0.2,0.15)$) {$\delta_{2}$};
            \node at ($(d3.east) + (0.2,0.15)$) {$\delta_{3}$};
            \node at ($(d4.east) + (0.2,0.15)$) {$\delta_{4}$};
            \node at ($(d5.east) + (0.2,0.15)$) {$\delta_{5}$};
            \node at ($(d6.east) + (0.2,0)$) {$\delta_{6}$};
            \node at ($(d0.east) + (0.2,0)$) {$\delta_{0}$};
	    \end{tikzpicture}} \qquad
		\begin{alignedat}{2}
			\delta_{0} &= \mathcal{E}_{7} - \mathcal{E}_{8} , &\qquad  \delta_{4} &= \mathcal{H}_{p} - \mathcal{E}_{3} - \mathcal{E}_{5}, \\
			\delta_{1} &= \mathcal{E}_{1} - \mathcal{E}_{2}, &\qquad  \delta_{5} &=  \mathcal{E}_{3} - \mathcal{E}_{4}, \\
			\delta_{2} &= \mathcal{H}_{q} - \mathcal{E}_{1} - \mathcal{E}_{5} , &\qquad  
						\delta_{6} &= \mathcal{E}_{6} - \mathcal{E}_{7}.\\
			\delta_{3} &= \mathcal{E}_{5} - \mathcal{E}_{6},
		\end{alignedat}
    \end{equation}
	\caption{The standard surface root basis for the $E_{6}^{(1)}$ surface} \label{fig:std-surf-roots}	
\end{figure}
The standard surface root basis is shown on Figure~\ref{fig:std-surf-roots}	 
and the standard symmetry root bases is shown on Figure~\ref{fig:std-symm-roots}.
\begin{figure}[ht]
    \begin{equation}\label{eq:std-symm-roots}			
	    \raisebox{-28.0pt}{\begin{tikzpicture}[scale=1.6, elt/.style={circle, draw=black!100, thick, inner sep=0pt, minimum size=2mm}]
            \path (-0.6,-0.3) node (a1) [elt] {}
            ( 0.6,-0.3) node (a2) [elt] {}
            ( 0,0.67) node (a0) [elt] {};
            \draw [black] (a1) -- (a2) -- (a0) -- (a1);
            \node at ($(a1.west) + (-0.2,0.0)$) {$\alpha_{1}$};
            \node at ($(a2.east) + (0.2,0.0)$) {$\alpha_{2}$};
            \node at ($(a0.north) + (0,0.2)$) {$\alpha_{0}$};
	    \end{tikzpicture}} \qquad
		\begin{aligned}
			\alpha_{0} &= \mathcal{H}_{q} + \mathcal{H}_{p} - \mathcal{E}_{5} - \mathcal{E}_{6} - \mathcal{E}_{7} - \mathcal{E}_{8}\\
			\alpha_{1} &= \mathcal{H}_{q} - \mathcal{E}_{3} - \mathcal{E}_{4}, \\
			\alpha_{2} &= \mathcal{H}_{p} - \mathcal{E}_{1} - \mathcal{E}_{2}.
		\end{aligned}
    \end{equation}
	\caption{The standard symmetry root basis for the $E_{6}^{(1)}$ surface} 
	\label{fig:std-symm-roots}
\end{figure}

\subsubsection{The affine Weyl symmetry group $\widetilde{W}(A_{2}^{(1)})$} 
\label{ssub:symm-group}
Recall that the algebraic origin of the Painlev\'e dynamic is the birational representation of the extended affine Weyl
symmetry group $\widetilde{W}\left(A_{2}^{(1)}\right) = \operatorname{Aut} \left(A^{(1)}_2\right) \ltimes W\left(A^{(1)}_2\right)$.
The affine Weyl group $W\left(A^{(1)}_2\right)$ is described in terms of the generators $w_i$ and relations encoded in the diagram 
on Figure~\ref{fig:std-symm-roots},
\begin{equation*}
W\left(A_{2}^{(1)}\right) =
\left\langle w_{0}, w_1, w_{2}\ \left|\ 
\begin{alignedat}{2}
    w_{i}^{2} = e, \quad  w_{i}\circ w_{j} &= w_{j}\circ w_{i}& 
    &\quad\text{ when \raisebox{-0.08in}{
    \begin{tikzpicture}[elt/.style={circle,draw=black!100,thick, inner sep=0pt,minimum size=1.5mm}]
        \path ( 0,0) node (ai) [elt] {} ( 0.5,0) node (aj) [elt] {};
        \draw [black] (ai)  (aj);
        \node at ($(ai.south) + (0,-0.2)$) 	{$\alpha_{i}$};
        \node at ($(aj.south) + (0,-0.2)$)  {$\alpha_{j}$};
    \end{tikzpicture}}}\\
    w_{i}\circ w_{j}\circ w_{i} &= w_{j}\circ w_{i}\circ w_{j}& &\quad\text{ when 
	\raisebox{-0.17in}{\begin{tikzpicture}[elt/.style={circle,draw=black!100,thick, inner sep=0pt,minimum size=1.5mm}]
        \path ( 0,0) node (ai) [elt] {} ( 0.5,0) 	node  	(aj) [elt] {};
        \draw [black] (ai) -- (aj);
        \node at ($(ai.south) + (0,-0.2)$) 	{$\alpha_{i}$};
        \node at ($(aj.south) + (0,-0.2)$)  {$\alpha_{j}$};
    \end{tikzpicture}}}
\end{alignedat}\right.\right\rangle. 
\end{equation*} 
and the group $\operatorname{Aut} \left(A^{(1)}_2\right)$ of the Dynkin diagram automorphisms is isomorphic to the 
dihedral group $\mathbb{D}_{3}$ of the symmetries of the triangle that is generated by two reflections $\sigma_{1}$
and $\sigma_{2}$ that act on the symmetry and the surface roots as permutations,
\begin{equation}\label{eq:A2-autos}
	\sigma_{1} = (\alpha_{1}\alpha_{2}) = (\delta_{1}\delta_{5})(\delta_{2}\delta_{4}),\qquad
	\sigma_{2} = (\alpha_{0}\alpha_{2}) = (\delta_{0}\delta_{1})(\delta_{2}\delta_{6}),	
\end{equation}
where we have used the standard cycle notation for the permutations. This group can be realized using reflections
on the Picard lattice, where $w_{i}$ are given in terms of reflections in the symmetry roots $\alpha_{i}$
\begin{align}\label{eq:Pic-action-w}
	r_{\alpha_{i}}(\mathcal{C}) &= \mathcal{C} + (\mathcal{C}\bullet \alpha_{i}) \alpha_{i}, 
	\qquad \mathcal{C}\in \operatorname{Pic}(\mathcal{X})\\
	\intertext{ and $\sigma_{i}$ can also be given in terms of reflections in other lattice vectors,}
	\sigma_{1} &= r_{\mathcal{H}_{q}-\mathcal{H}_p}\circ r_{\mathcal{E}_{1} - \mathcal{E}_{3}}\circ r_{\mathcal{E}_{2} - \mathcal{E}_{4}},
	\quad
	\sigma_{2} = r_{\mathcal{E}_{1} - \mathcal{E}_{7}} 
	\circ r_{\mathcal{E}_{2} - \mathcal{E}_{8}} \circ r_{\mathcal{H}_{q}-\mathcal{E}_{5} - \mathcal{E}_{6}}.\label{eq:Pic-action-auto}
\end{align}


\begin{lemma}\label{lem:W2-bir} The birational representation of $\widetilde{W}(A^{(1)}_2)$ 
is given by the following birational maps of $\mathbb{P}^{1} \times \mathbb{P}^{1}$, 
that we also denote by $w_{i}$ and $\sigma_{j}$, that lift to the isomorphisms of the surfaces $\mathcal{X}_{\mathbf{a}}$ (note
that the root variables and hence the location of the blowup points change, and that change is consistent with the action of 
the group on the roots) and whose induced action on the Picard lattice is given by \eqref{eq:Pic-action-w}--\eqref{eq:Pic-action-auto}:
\begin{align*}
		w_{0}: 
		\left(\begin{matrix} a_{0} \quad a_{1} \\ a_{2} \end{matrix}; \ t \ ;
		\begin{matrix} f \\ g \end{matrix}\right) 
		&\mapsto 
		\left(\begin{matrix} -a_{0} \quad a_{1} + a_{0} \\ a_{2} + a_{0} \end{matrix} ; \ t \ ;
		\begin{matrix}  f + \frac{a_{0}}{g - f - t}  \\[3pt]  g + \frac{a_{0}}{g - f - t}  \end{matrix}\right),
		 \\
		w_{1}: \left(\begin{matrix}  a_{0} \quad a_{1} \\ a_{2} \end{matrix} ;\ t\ ;
		\begin{matrix} f \\ g \end{matrix}\right)
		&\mapsto 
		\left(\begin{matrix} a_{0} + a_{1} \quad -a_{1} \\ a_{2} + a_{1} \end{matrix} ; \ t \ ;
		\begin{matrix}  f \\  g - \frac{a_{1}}{f} \end{matrix}\right),
		\\
		w_{2}: 
		\left(\begin{matrix}  a_{0} \quad a_{1} \\ a_{2} \end{matrix} ;\ t\ ;
		\begin{matrix} f \\ g \end{matrix}\right) 
		&\mapsto 
		\left(\begin{matrix} a_{0} + a_{2} \quad a_{1} + a_{2} \\ -a_{2} \end{matrix} ;\ t\ ;
		\begin{matrix}  f + \frac{a_{2}}{g}\\ g \end{matrix}\right),\\
		\sigma_{1}: 
		\left(\begin{matrix} a_{0} \quad a_{1} \\ a_{2} \end{matrix}\ ;\ t\ ;
		\begin{matrix} f \\ g \end{matrix}\right) 
		&\mapsto 
		\left(\begin{matrix} -a_{0} \quad -a_{2} \\ -a_{1} \end{matrix}\ ;\ t\ ;
		\begin{matrix}  -g \\ -f \end{matrix}\right),
        \\
		\sigma_{2}: \left(\begin{matrix}  a_{0} \quad a_{1} \\ a_{2} \end{matrix}\ ;\ t\ ;
		\begin{matrix} f \\ g \end{matrix}\right)
		&\mapsto 
		\left(\begin{matrix} -a_{2} \quad -a_{0} \\ -a_{1} \end{matrix}\ ;\ t\ ;
		\begin{matrix}  g - f - t \\  - f \end{matrix}\right).			
\end{align*}
\end{lemma}
The proof of this Lemma follows the same lines as in
\cite{DzhFilSto:2020:RCDOPWHWDPE} and is omitted; see also \cite{KajNouYam:2017:GAPE}.

\subsubsection{Standard discrete Painlev\'e equation on the $E_{6}^{(1)}$ surface} 
\label{ssub:std-dP-E6}
The standard discrete Painlev\'e equation $[01\overline{1}]$ (in notation of \cite{LiDzhFilZha:2025:RRGLCOPDPDS})
corresponds to the dynamics acting on the root variables $a_{i}$ (equivalently, on the symmetry roots $\alpha_{i}$) as
\begin{equation}\label{eq:std-evol}
	\overline{a}_{0}=a_{0},\quad \overline{a}_{1}=a_{1}-1,\quad \overline{a}_{2}=a_{2}+1;\qquad
	\overline{\mathbf{\upalpha}}=\mathbf{\upalpha} + \langle 0,1,-1 \rangle \delta
\end{equation}
where $\mathbf{\upalpha} = \langle\alpha_{0},\alpha_{1},\alpha_{2}\rangle$ and 
$\delta = -\mathcal{K}_{\mathcal{X}} = \alpha_{0} + \alpha_{1} + \alpha_{2}$. This action can be represented
in terms of generators as $\psi = \sigma_{1}\circ \sigma_{2}\circ w_{0}\circ w_{2}$, which results in the equations
\eqref{eq:dP-std},
\begin{equation*}
			\overline{q} + q = p - t - \frac{a_2}{p},\quad
			p + \underline{p} = q + t + \frac{a_1}{q}.	
\end{equation*}

\subsection{Matching the Min-Chen and the standard dynamic} 
\label{sub:match}
	To match the dynamics \eqref{eq:MinChen} and \eqref{eq:dP-std}, we first need to match the geometries by finding the change of
	basis of the Picard lattice identifying the surface roots on Figures~\ref{fig:MC-surf-roots} and \ref{fig:std-surf-roots}.
	It is quite straightforward to see that such a change is given by
	\begin{alignat*}{2}\label{eq:basis-change}
	    \mathcal{H}_{x}& = 2\mathcal{H}_{q} + \mathcal{H}_{p} - \mathcal{E}_{3} -\mathcal{E}_{5}-\mathcal{E}_{6}-\mathcal{E}_{7}, \qquad& 
			\mathcal{H}_{q} &= \mathcal{H}_{x} + \mathcal{H}_{y} - \mathcal{F}_{1} - \mathcal{F}_{4},  \\
	    \mathcal{H}_{y}& = \mathcal{H}_{q} +\mathcal{H}_{p} - \mathcal{E}_{5} - \mathcal{E}_{6}, \qquad&  
			\mathcal{H}_{p} &= \mathcal{H}_{x} + 2\mathcal{H}_{y} - \mathcal{F}_{1}- \mathcal{F}_{4}- \mathcal{F}_{5}- \mathcal{F}_{6}, \\
	    \mathcal{F}_{1} &= \mathcal{H}_{q} +\mathcal{H}_{p} - \mathcal{E}_{5} - \mathcal{E}_{6} - \mathcal{E}_{7}, \qquad& 
			\mathcal{E}_{1} &= \mathcal{F}_{7}, \\
	    \mathcal{F}_{2} &= \mathcal{E}_{8}, \qquad& \mathcal{E}_{2} &= \mathcal{F}_{8}, \\
		\mathcal{F}_{3} &= \mathcal{E}_{4}, \qquad& \mathcal{E}_{3} &= \mathcal{H}_{y} - \mathcal{F}_{4}, \\
		\mathcal{F}_{4} &= \mathcal{H}_{q} +\mathcal{H}_{p} - \mathcal{E}_{3} - \mathcal{E}_{5}- \mathcal{E}_{6}, \qquad& 	
			\mathcal{E}_{4} &= \mathcal{F}_{3}, \\
		\mathcal{F}_{5} &= \mathcal{H}_{q} - \mathcal{E}_{6}, \qquad& 
			\mathcal{E}_{5} &= \mathcal{H}_{x} + \mathcal{H}_{y} - \mathcal{F}_{1} - \mathcal{F}_{4}- \mathcal{F}_{6}, \\
		\mathcal{F}_{6} &= \mathcal{H}_{q} - \mathcal{E}_{5}, \qquad& 
			\mathcal{E}_{6} &= \mathcal{H}_{x} + \mathcal{H}_{y} - \mathcal{F}_{1} - \mathcal{F}_{4}- \mathcal{F}_{5}, \\
		\mathcal{F}_{7} &= \mathcal{E}_{1}, \qquad& \mathcal{E}_{7} &= \mathcal{H}_{y} - \mathcal{F}_{1}, \\
		\mathcal{F}_{8} &= \mathcal{E}_{2}, \qquad& \mathcal{E}_{8} &= \mathcal{F}_{2}.
	\end{alignat*}	
This gives the expressions for the symmetry roots of the Chen-Min recurrence,
\begin{equation*}
	\alpha_{0} = \mathcal{F}_{1} - \mathcal{F}_{2},\quad \alpha_{2}=\mathcal{H}_{x} - \mathcal{F}_{1} - \mathcal{F}_{3},\quad 
	\alpha_{3}=\mathcal{H}_{x} + 2 \mathcal{H}_{y} - \mathcal{F}_{1} - \mathcal{F}_{4}-\mathcal{F}_{5}-\mathcal{F}_{6}
	-\mathcal{F}_{7}-\mathcal{F}_{8},
\end{equation*}
and applying the evolution \eqref{eq:MinChen-evol} we see that the symmetry roots evolve as
\begin{equation*}
	\alpha_{0}\mapsto \alpha_{0},\ \alpha_{1} \mapsto \alpha_{1} + \delta,\ \alpha_{2}\mapsto \alpha_{2}- \delta,\quad
	\delta = 2 \mathcal{H}_{x} + 2 \mathcal{H}_{y} - \mathcal{F}_{1}-\cdots-\mathcal{F}_{8}=-\mathcal{K}_{\mathcal{X}},
\end{equation*}
so we indeed get the standard dynamic. Moreover, using the \emph{period map} computation, as in \cite{DzhFilSto:2020:RCDOPWHWDPE},
we get the expressions for the root variables in terms of the step $n$, 
\begin{equation}\label{eq:MinChen-rvs}
	a_{0}=0,\quad a_{1}=-n,\quad a_{2}=n+1
\end{equation}
whose evolution with the step $n\mapsto n+1$ again coincides with the standard dynamic \eqref{eq:std-evol}. However, note that 
the symmetry root $\alpha_{0} = \mathcal{F}_{1} - \mathcal{F}_{2}$ in now the class \emph{effective divisor} of the nodal curve
$F_{1} - F_{2}$ on Figure~\ref{fig:MC-surf}, and that is the reason the corresponding root variable vanishes. This completes the proof
of Theorem~\ref{eq:math-thm}.

\section{Comparison with the modified Laguerre case} 
\label{sec:mod-Laguerre}
There are other examples of orthogonal polynomials that are related to discrete Painlev\'e equations on the $E_{6}^{(1)}$ surface.
Indeed, one such example is the recurrence coefficients for the orthogonal polynomials with the modified Laguerre weight 
$\tilde{w}(x;\alpha) = x^{\alpha} e^{-x^{2} + \tau x}$ studied by L.~Boelen and W.~Van Assche \cite{BoeVan:2010:DPERCSLP}, 
G.~Filipuk, W.~Van Assche, and L.~Zhang \cite{FilVanZha:2012:RCSLPFPE}, and 
P.~Clarkson and K.~Jordaan \cite{ClaJor:2014:RBSLPFPE}; see also \cite{Van:2018:OPPE} and \cite{DzhFilSto:2022:DERCSOPTRPEGA}.
The recurrence coefficients for the monic polynomials with this weight can be parameterized using variables $(f_{n},g_{n})$ 
satisfying the recurrence relation
\begin{equation}\label{eq:modLag}
			f_{n} f_{n-1} = \frac{g_{n} + n + \frac{\alpha}{2}}{g_{n}^{2} - \frac{\alpha^{2}}{4}},\quad
			g_{n} + g_{n+1} = \frac{1}{f_{n}}\left(\frac{\tau}{\sqrt{2}} - \frac{1}{f_{n}}\right)
\end{equation}
The recurrence \eqref{eq:MinChen} then can be easily seen as a limiting case $\alpha\to0$ of \eqref{eq:modLag} with the coordinate
identification with $x=g$, $y=\sqrt{2}/f$, and $s=\tau/2$. Under this limit the modified Laguerre weight becomes re-centered Gaussian
weight $\tilde{w}(a;0) = e^{-x^{2} + \tau x} = e^{-(x-s)^{2} + s^{2}}$ where the shift corresponds to the location of the jump of the 
weight $ w(x, s) = e^{-x^{2}} (A + B\, \theta(x - s))$ that we consider. Geometrically we see how such a limit results in the 
appearance of the nodal curve. The Sakai surface for the modified Laguerre weight is shown on Figure~\ref{fig:modLag-surf}
\begin{center}
\begin{figure}[ht]
        \centering
	\begin{tikzpicture}[scale=0.9, >=stealth,basept/.style={circle, draw=red!100, fill=red!100, thick, inner sep=0pt,minimum size=1.2mm}, scale=1.3]
	\begin{scope}[xshift=0cm,yshift=0cm]
	\draw [black, line width = 1pt] (-0.4,0.5) -- (2.9,0.5)	node [pos=0,left] {\small $H_{g}$} node [pos=1,right] {\small $g=0$};
	\draw [black, line width = 1pt] (-0.4,2.5) -- (2.9,2.5) node [pos=0,left] {\small $H_{g}$} node [pos=1,right] {\small $g=\infty$};
	\draw [black, line width = 1pt] (0,-0.4) -- (0,2.9) node [pos=0,below] {\small $H_{f}$} node [pos=1,yshift=3pt] {\small $f=0$};
	\draw [black, line width = 1pt] (2.5,-0.4) -- (2.5,2.9) node [pos=0,below] {\small $H_{f}$} node [pos=1,yshift=3pt] {\small $f=\infty$};
	\node (w1) at (2.5,0.1) [basept,label={[xshift = 8pt, yshift=-8pt] \small $w_{1}$}] {};
	\node (w2) at (2.5,0.9) [basept,label={[xshift = 8pt, yshift=-8pt] \small $w_{2}$}] {};
	\node (w3) at (0,1.7) [basept,label={[xshift=-8pt,yshift=-8pt] \small $w_{3}$}] {};
	\node (w4) at (0,2.5) [basept,label={[xshift = -8pt, yshift=-15pt] \small $w_{4}$}] {};
	\node (w5) at (1,3.5) [basept,label={[above] \small $w_{5}$}] {};
	\node (w6) at (1.7,3.5) [basept,label={[above] \small $w_{6}$}] {};
	\node (w7) at (2.4,3.5) [basept,label={[above] \small $w_{7}$}] {};
	\node (w8) at (3.1,3.5) [basept,label={[above] \small $w_{8}$}] {};
	\draw [red, line width = 0.8pt, ->] (w5) -- (0.5,2.5) -- (w4);
	\draw [red, line width = 0.8pt, ->] (w6) -- (w5);
	\draw [red, line width = 0.8pt, ->] (w7) -- (w6);
	\draw [red, line width = 0.8pt, ->] (w8) -- (w7);
	\end{scope}
	\draw [->] (5,1.7)--(3,1.7) node[pos=0.5, below] {$\operatorname{Bl}_{w_{1}\cdots w_{8}}$};
	\begin{scope}[xshift=6.3cm,yshift=0cm]
	\draw [black, line width = 1pt] (-0.4,0.5) -- (3.5,0.5) node [pos=1, right] {\small $g=0$};
	\draw [blue, line width = 1pt] (3,-0.4) -- (3,2.9) node [pos=0, below] {\small $H_{f}-K_{1}-K_{2}$};
	\draw [red, line width = 1pt] (2.5,0.3) -- (3.2,-0.4) node [pos=1, xshift = 6pt, yshift=4pt] {\small $K_{1}$};
	\draw [red, line width = 1pt] (-0.4,0.1) -- (2.9,0.1) node [pos=0, left] {\small $H_{g}-K_{1}$};
	\draw [red, line width = 1pt] (2.5,0.7) -- (3.2,1.4) node [pos=1, xshift = 6pt, yshift=-4pt] {\small $K_{2}$};
	\draw [red, line width = 1pt] (-0.4,0.9) -- (2.9,0.9) node [pos=0, left] {\small $H_{g}-K_{2}$};
	\draw [blue, line width = 1pt] (0,-0.4) -- (0,2.9) node [pos=0, below] {\small $H_{f}-K_{3}-K_{4}$};
	\draw [red, line width = 1pt] (-0.2,1.2) -- (0.5,1.9) node [pos=0, xshift = -6pt, yshift=4pt] {\small $K_{3}$};
	\draw [red, line width = 1pt] (0.1,1.7) -- (3.4,1.7) node [pos=1, right] {\small $H_{g}-K_{3}$};
	\draw [blue, line width = 1pt] (-0.2,2.3) -- (0.8,3.3) node [pos=0, xshift = -17pt, yshift=-5pt] {\small $K_{4}-K_{5}$};
	\draw [blue, line width = 1pt] (0.3,3.3) -- (1.3,2.3) node [pos=1, xshift = 17pt, yshift=-5pt] {\small $K_{5}-K_{6}$};
	\draw [blue, line width = 1pt] (0.9,2.5) -- (3.4,2.5) node [pos=1, right] {\small $H_{g}-K_{4}-K_{5}$};
	\draw [blue, line width = 1pt] (0.6,2.7) -- (1.6,3.7) node [pos=1, right, yshift=2pt] {\small $K_{6}-K_{7}$};
	\draw [blue, line width = 1pt] (0.8,3.7) -- (1.8,2.7) node [pos=0, left, yshift=2pt] {\small $K_{7}-K_{8}$};
	\draw [red, line width = 1pt] (1.4,2.8) -- (2.1,3.5) node [pos=1, xshift = 6pt, yshift=-4pt] {\small $K_{8}$};
	\end{scope}
	\end{tikzpicture}
	\caption{The Sakai surface for the modified Laguerre weight recurrence relation}
	\label{fig:modLag-surf}
\end{figure}	
\end{center}
where the coordinates of the base points are 
\begin{equation*}
	w_{1}\left(F=\frac{1}{f}=0,g=-\frac{\alpha}{2}\right),\quad w_{2}\left(F=0,g=\frac{\alpha}{2}\right),\quad 
		w_{3}\left(f=0,g=-n-\frac{\alpha}{2}\right)
\end{equation*}
and the cascade
\begin{align*}
	w_{4}\left(f=0,G=\frac{1}{g}=0\right)&\leftarrow w_{5}\left(u_{4}=f=0,v_{4}=\frac{f}{G}=0\right)\\
	&\leftarrow w_{6}\left(u_{5}=u_{4}=0,v_{5}=\frac{v_{4}}{u_{4}}=-1\right)\\
	&\leftarrow w_{7}\left(u_{6}=u_{5}=0,v_{6}=\frac{v_{5}+1}{u_{5}}=-\frac{\tau}{\sqrt{2}}\right)\\
	&\leftarrow w_{8}\left(u_{7}=u_{6}=0,v_{7}=\frac{2v_{6}+\sqrt{2}\tau}{2u_{6}}=-1-n-\frac{\alpha+\tau^{2}}{2}\right).
\end{align*}
Thus, this recurrence correspond to the generic point configuration with the full symmetry group $\widetilde{W}\left(A_{2}^{(1)}\right)$.
This surface on Figure~\ref{fig:modLag-surf} is, up to some rotation of the coordinate axes, essentially the same as Figure~\ref{fig:MC-surf},
with one major difference -- the points $w_{1}$ and $w_{2}$ are distinct. As we take the limit $\alpha\to0$,
the points $w_{1}$ and $w_{2}$ collide and that results in the appearance of the nodal curve on Figure~\ref{fig:MC-surf}.

\begin{acknowledgement}
The authors would like to thank Galina Filipuk, Alexander Stokes, and Ralph Willox for many helpful discussions. We also thank the referee for valuable feedback. 
\end{acknowledgement}

\bibliographystyle{amsalpha}

\providecommand{\bysame}{\leavevmode\hbox to3em{\hrulefill}\thinspace}
\providecommand{\MR}{\relax\ifhmode\unskip\space\fi MR }
\providecommand{\MRhref}[2]{%
  \href{http://www.ams.org/mathscinet-getitem?mr=#1}{#2}
}
\providecommand{\href}[2]{#2}

\end{document}